\documentclass[11pt,twoside]{article}
\usepackage{asp2006}
\usepackage{psfig}
\usepackage{epsf}
\usepackage{graphics}
\usepackage{lscape}
\def\edcomment#1{\iffalse\marginpar{\raggedright\sl#1\/}\else\relax\fi}
\reversemarginpar

\begin{document}
\title{VLA survey of the CDFS: the nature of faint radio sources}

\author{P. Tozzi} 
\affil{INAF, Osservatorio Astronomico di Trieste, via G.B. Tiepolo 11, I-34143 Trieste, Italy}
\author{K. Kellermann, E. Fomalont} 
\affil{NRAO, 520 Edgemont Road, Charlottesville, VA~22903--2475, U.S.A.} 
\author{N. Miller, C. Norman}
\affil{Johns Hopkins University,
  3400 North Charles Street, Baltimore, MD 21218, USA}
\author{V. Mainieri, P. Padovani, P. Rosati} \affil{ESO,
  Karl-Schwarschild-Strasse 2, D-85748} \author{and the VLA--CDFS
  team}

\begin{abstract}
We present the multiwavelength properties of 266 cataloged radio
sources identified with 20 and 6 cm VLA deep observations of the
CDFS at a flux density limit of $42~\mu$Jy at the field centre at
1.4 GHz.  These new observations probe the faint end of both the star
formation and radio galaxy/AGN population.  X--ray data, including
upper limits, turn out to be a key factor in establishing the nature
of faint radio sources.  We find that, while the well--known
flattening of the radio number counts below 1 mJy is mostly due to
star forming galaxies, these sources and AGN make up an approximately
equal fraction of the sub--millijansky sky, contrary to some previous
results.  We have also uncovered a population of distant AGN
systematically missing from many previous studies of sub--millijansky
radio source identifications. The AGN include radio galaxies, mostly
of the low--power, Fanaroff--Riley I type, and a significant
radio--quiet component, which amounts to approximately one fifth of the
total sample.  We also find that radio detected, X--ray AGN are not
more heavily obscured than the X--ray detected AGN.  This argues
against the use of radio surveys as an efficient way to search for the
missing population of strongly absorbed AGN.

\end{abstract}

\vspace{-0.5cm}
\section{Introduction}

Deep multiwavelength surveys help to reconstruct the cosmic evolution
of AGN and star formation processes.  In this respect, X--ray and
radio emission are good tracers of both processes.  The radio
properties of the X--ray population found in deep surveys have been
studied only in a few papers based on VLA data in the Chandra Deep
Field North (CDFN; Richards et al. 1998; Richards 2000; Bauer et
al. 2002; Barger et al. 2007), combined MERLIN and VLA data in the
CDFN region (Muxlow et al. 2005), and ATCA data in the Chandra Deep
Field South (CDFS; Afonso et al. 2006; Rovilos et al. 2007).  Deep
radio surveys are also realized in shallower but wider X--ray fields
like COSMOS (see Schinnerer et al. 2007; Smolcic et al. 2008; 2009).

In this work, we use the deep radio data obtained with the VLA in the
CDFS and Extended Chandra Deep Field South (E-CDFS) fields.  The radio
catalog (presented in Kellermann et al. 2008, hereafter Paper I)
includes 266 sources and constitutes one of the largest and most
complete samples of $\mu$Jy sources in terms of redshift information.
Our multiwavelength approach exploits the X--ray data (see Giacconi et
al. 2002; Alexander et al. 2003; Lehmer et al. 2005) and allows us to
characterize both processes over a wide range of redshifts.  Optical
and near--IR properties of the radio sources are discussed by Mainieri
et al. (2008, hereafter Paper II), a detailed analysis of the X--ray
properties of radio sources is presented in Tozzi et al. 2009
(hereafter Paper III), while a multiwavelength approach to studying
the source population is presented by Padovani et al. (2009, Paper
IV).  Here we briefly discuss the most relevant outcomes presented in
this series of papers.  


\section{The data set}
 
Of 266 cataloged radio sources, 89 radio sources in our complete radio
catalog were found to have X--ray counterparts in either the 1
Megasecond Chandra catalog or in the E-CDFS. Using the available
imaging in {\it i}, R, K$_{\rm S}$, 3.6 $\mu m$, 4.5 $\mu m$, 5.8 $\mu
m$, 8.5 $\mu m$, 24 $\mu m$ and 70 $\mu m$ bands from ESO/WFI,
VLT/ISAAC, HST/ACS and Spitzer, we were able to find a reliable
optical or infrared counterpart for 254 ($\sim 95 \%$) radio
sources. Three radio sources have no apparent counterpart at any other
wavelength.  We also have optical morphological classifications for
$\sim$ 61\% of the sample.

Using literature data and our own follow--up, a total of 186
($\sim70\%$) sources have a redshift: 108 are spectroscopic redshifts
and 78 reliable photometric redshifts.  The redshift distribution of
the VLA sources peaks around $z \approx 0.8$ (see Figure
\ref{redshift}, left panel). The radio sources are good tracers of
large scale structures already detected at other wavebands in this
region of the sky (NIR, optical, X--ray). In particular, the main peaks
of our redshifts distribution are at z$=0.735 \pm 0.004$ (10 objects)
and z$=1.614 \pm 0.011$ (6 objects), two well known overdensities in
the CDFS (Gilli et al. 2003).

\section{Multiwavelength Classification of sub--mJy  radio sources}

\begin{figure}
\plottwo{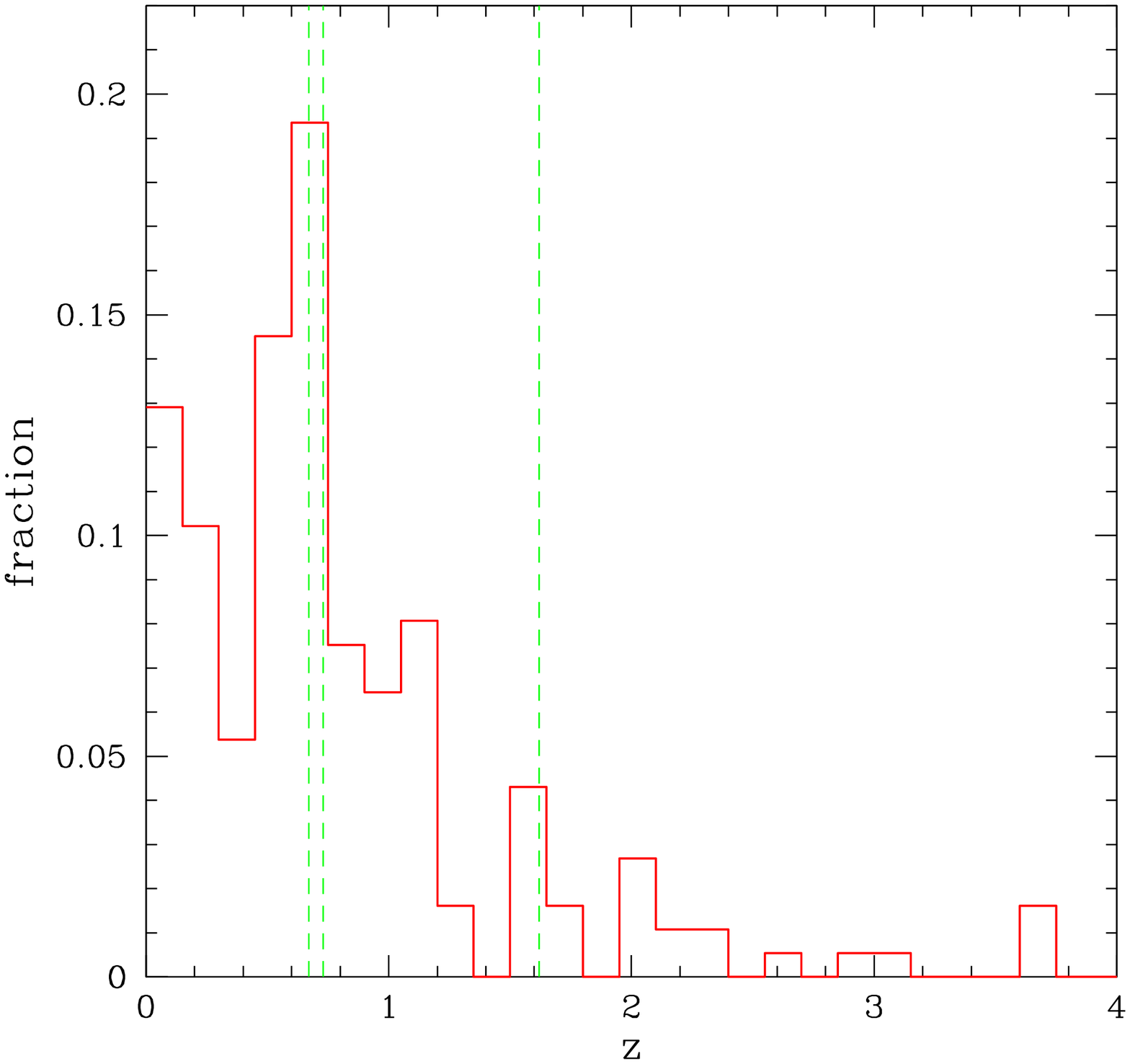}{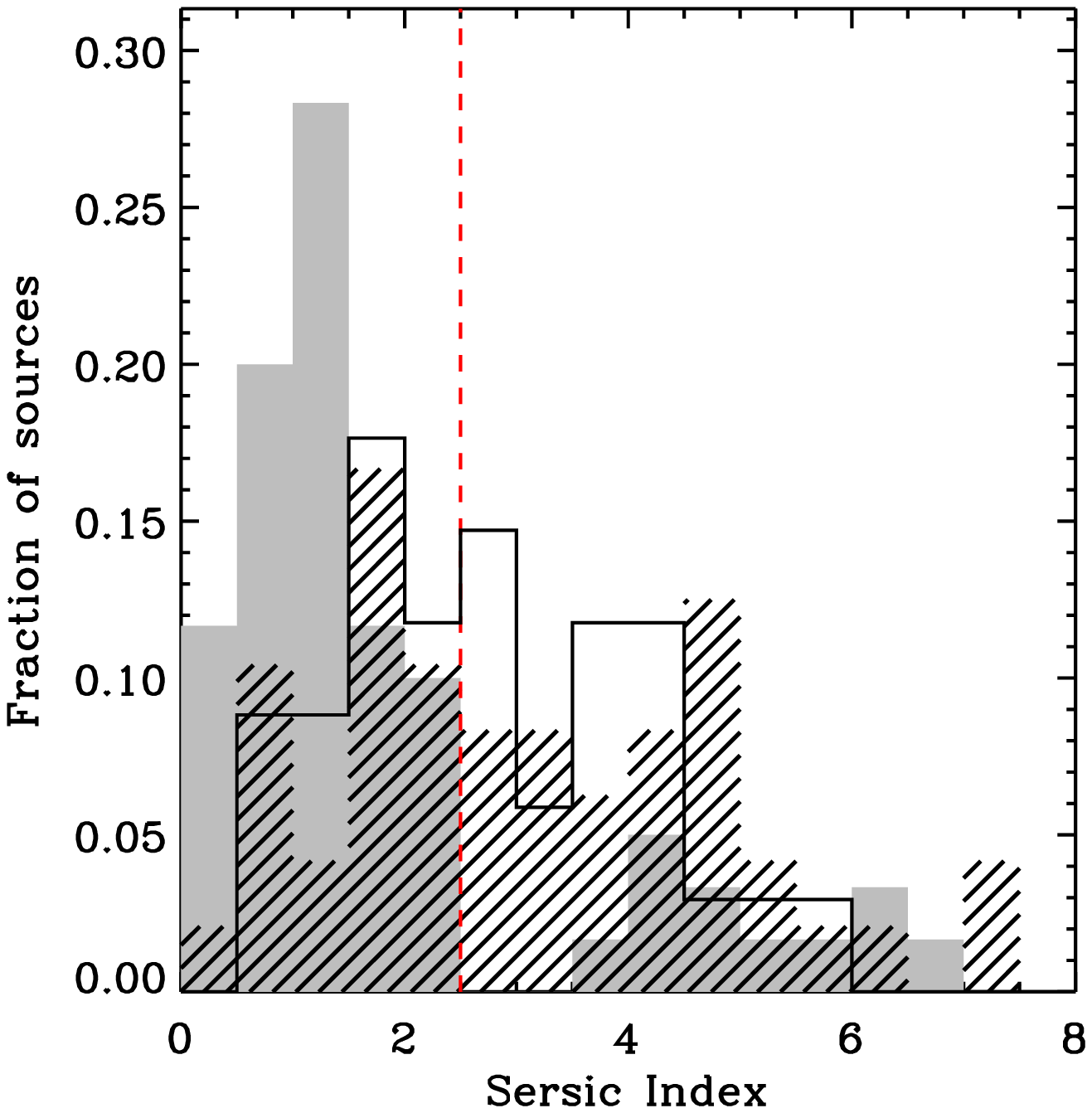}
\caption{Left panel: the fractional distribution of the 186 radio
  sources with redshift.  Dashed vertical lines mark three
  overdensities already identified in the X--ray sources distribution
  (Gilli et al. 2003).  Right panel: distribution of the Sersic index
  values for radio sources with S(1.4 GHz)$>0.2$ mJy (continuous
  line), $0.08<$S(1.4 GHz)$<0.2$ (hatched histogram) and S(1.4
  GHz)$<0.08$ mJy (grey shaded histogram). The vertical line mark the
  value $n=2.5$, empirical dividing value between early and late type
  galaxies. }
\label{redshift}
\end{figure}

By analyzing the ratio of radio to optical luminosity and the radio
and X--ray powers of the sources with morphological and redshift
information, we have selected candidate star--forming galaxies and AGN.
As the first two parameters by themselves are not very good
discriminants between star--forming galaxies and AGN, optical
morphology and especially X--ray data turned out to be vital in
establishing the nature of faint radio sources. 

In Figure \ref{redshift}, right panel, we show the distribution of the
Sersic index (see Sersic 1968) values for radio sources in three
equally populated flux density bins (see Paper II).  While the
properties of the host galaxies in the two brighter flux density bins
look similar, we find evidences for a change in the dominant radio
population at S$\approx 0.08$ mJy. The radio sources in the
intermediate and bright flux density bins show a Sersic indexes
distribution that resembles that of early--type galaxies with a tail
of disk dominated galaxies and in a rest--frame color--magnitude
diagram (U$-V$ versus M$_{\rm V}$) are preferentially ($70\%$) located
between the early--type/red--sequence galaxies.  On the other hand,
sources with S$< 0.08$ mJy have a Sersic indexes distribution that
peaks at low values of $n$, indicating a low value for the bulge to
disk ratio, with only $\approx 18 \%$ of the sources with $n>2.5$, and
they are widely spread in the color--magnitude diagram, with $\approx
60\%$ of them not being an early--type/red--sequence galaxy.

In Figure \ref{nh}, left panel, we show the fractional distribution of
measured intrinsic absorbing columns of equivalent $N_H$ for sources
with X--ray counterpart in the AGN luminosity range $L_{2-10} >
10^{42}$ erg s$^{-1}$ (see Paper III).  We find that in this
luminosity range $\sim 1/3$ of the sources are radio loud and $\sim
2/3$ radio quiet, where radio loud is defined as $log(R_X)>-2.9$ (with
$R_X \equiv \nu L_R(5 GHz) /L_{2-10}$).  We also find find a weak
anticorrelation of radio loudness as a function of intrinsic
absorption, adding support to the finding that radio emission is not
efficient in selecting more absorbed AGN.

\begin{figure}[!h]
 \plottwo{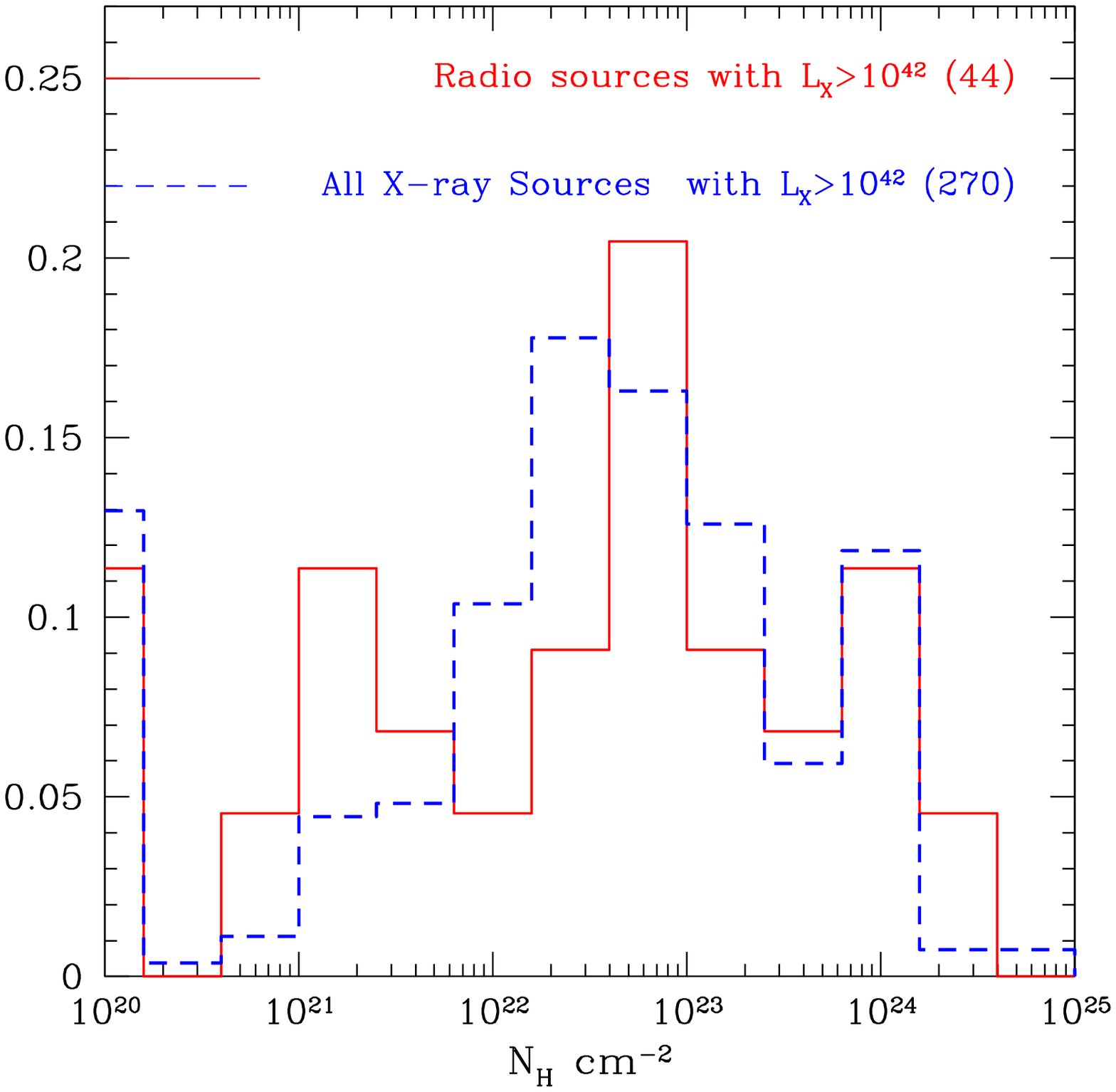}{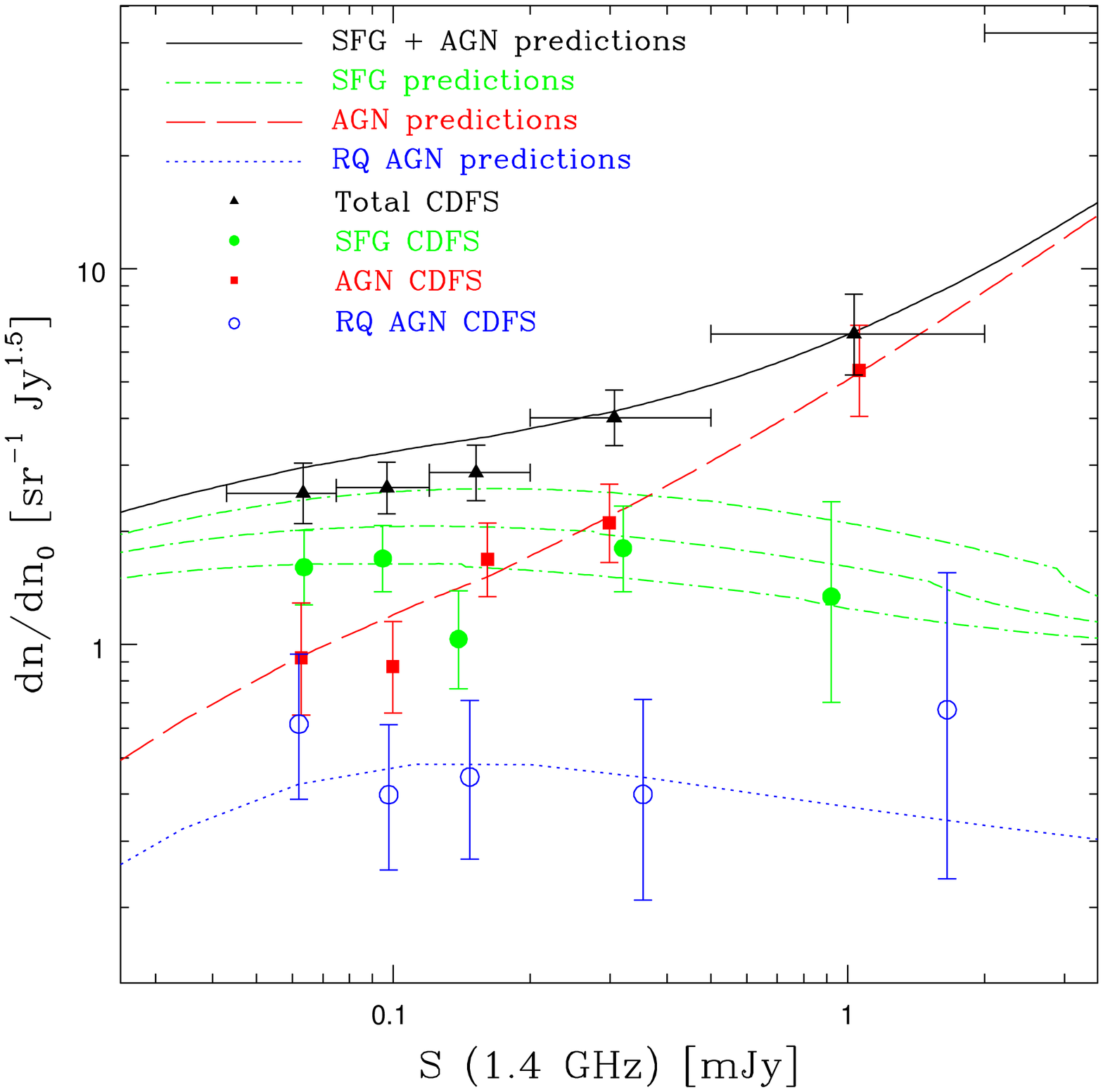}
 \caption{Left panel: fractional distribution of measured intrinsic
  absorbing columns of equivalent $N_H$ for sources with redshifts and
  $L_X >10^{42}$ erg s$^{-1}$ (continuous histogram). The fractional
  distribution of absorbing columns of the entire X--ray sample is
  also shown (dashed histogram).  The two distributions are consistent
  with each other.  Right panel: the Euclidean normalized 1.4 GHz
   CDFS source counts: total counts (black triangles), SFG (filled
   green circles), all AGN (red squares), and radio-quiet AGN (open
   blue circles). Error bars correspond to $1\sigma$ errors. Model
   calculations refer to SFG (green dot-dashed lines), displayed with
   a $1\sigma$ range on the evolutionary parameters, all AGN (red
   dashed line), radio-quiet AGN (blue dotted line), and the sum of
   the first two (black solid line). See Paper IV for more
   details.} \label{counts}
 \end{figure}

In Paper IV we exploit our multiwavelength approach to resolve the
faint radio number counts population.  As shown in Figure
\ref{counts}, right panel, we find that the well--known flattening of
the radio number counts below $\approx 1$ mJy is mostly due to
star--forming galaxies, which are missing above $\sim 2$ mJy but become
the dominant population below $\approx 0.1$ mJy.  AGN exhibit the
opposite behavior, as their counts drop at lower flux densities, going
from 100\% of the total at $\sim 10$ mJy down to $< 50\%$ at the
survey limit. This is driven by the fall of radio--loud sources, as
radio--quiet objects, which make up $\sim 20\%$ of sub--mJy sources,
display relatively flat counts.  Radio--quiet AGN make up about half of
all AGN. Their counts appear to be a scaled down version, by a factor
$\approx 3 - 4$, of those of star--forming galaxies, and are very
different from those of the radio--loud AGN population. This should
provide a clue to the origin of their radio emission.  

Star--forming galaxies make up $\la 60\%$ of sub--mJy sources down to
the flux limit of this survey. This has to be regarded as a robust
upper limit, as whenever we had to make some assumptions, we choose to
maximize their numbers. This result is at variance with the many
papers, which over the years have suggested a much larger dominance of
star forming galaxies below 1 mJy.  On the other hand, our results are
in broad agreement with a number of recent papers, which found a
significant AGN component down to $\approx 50~\mu$Jy.  The results of
our model calculations (see Paper IV for details) agree quite well
with the observed number counts and provide supporting evidence for
the scenario described above. Moreover, they imply that sub--mJy
radio--loud AGN are dominated by low--power, Fanaroff--Riley type~I
radio galaxies, as their high--power counterparts and radio--loud
quasars are expected to disappear below $\sim 0.5 - 1$ mJy.

\section{Conclusions}

We have used a deep radio sample, which includes 266 objects down to a
1.4 GHz flux density of $42~\mu$Jy selected in the Chandra Deep Field
South area, to study the nature of sub--mJy sources. Our unique set of
ancillary data, which includes reliable optical/near-IR
identifications, optical morphological classification, redshift
information, and X-ray detections or upper limits for a large fraction
of our sources, has allowed us to shed new light on this long standing
astrophysical problem.

Summarizing, we suggest that the flux density bin S $\ge 0.08$ mJy is
dominated by a population of early-type galaxies harboring low
luminosity AGN, while only at flux densities below $\approx 0.08$ mJy
starburst galaxies start to become dominant.  Considering the apparent
emerging population of low luminosity AGN at microjansky levels, care
is needed when interpreting radio source counts in terms of the
evolution of the star formation rate in the Universe.

We plan to expand on this work by using our deeper ($7~\mu$Jy per beam
over the whole Extended CDFS region) radio observations (Miller et
al. 2008) and the recently released 2 Msec Chandra data (Luo et
al. 2008), to obtain additional contraints on the role of star forming
galaxies as opposed to AGN in the sub--mJy radio population.

\begin{acknowledgements}
The VLA is a facility of the National Science Foundation operated by
NRAO under a cooperative agreement with Associated Universities Inc.
P. Tozzi acknowledges financial contribution from contract ASI--INAF
I/023/05/0 and from the PD51 INFN grant.
\end{acknowledgements}

\end{document}